\documentclass[aps,prb,twocolumn,amsmath,amssymb,preprintnumbers,floatfix]{revtex4}

\usepackage[utf8]{inputenc}
\usepackage{txfonts}
\usepackage{microtype}
\usepackage{eucal}
\usepackage{bm} 

\usepackage{color}
\usepackage{soul}

\usepackage{graphicx}
\usepackage{float}

\usepackage[colorlinks,allcolors=blue]{hyperref}
\usepackage[capitalize]{cleveref}

\renewcommand{\i}{\mathrm{i}}

\newcommand{\eg}{\textit{e.g.}\ }

\definecolor{Jacob}{rgb}{1,0,0}
\definecolor{Takehito}{rgb}{0,0,1}

\newcommand{\ve}[1]{\boldsymbol{#1}}

\newcommand{\vk}{{\ve{k}}} % Vector k

\newcommand{\e}[1]{\mathrm{e}^{#1}}
\def\i{\mathrm{i}}

\begin{document}
\title{Anisotropic Andreev Reflection and Josephson Effect in Ballistic Phosphorene}
\author{Jacob Linder}
\affiliation{Department of Physics, NTNU, Norwegian University of Science and Technology, N-7491 Trondheim, Norway}

\author{Takehito Yokoyama}
\affiliation{Department of Physics, Tokyo Institute of Technology, Tokyo 152-8551, Japan
}
\date{\today}
 
\begin{abstract}
  \noindent
We study Andreev reflection and the Josephson effect in a ballistic monolayer of black phosphorous, known as phosphorene. Due to the anisotropic band structure of this system, the supercurrent changes with an order of magnitude when comparing tunneling along two perpendicular directions in the monolayer. We show that the main reason for this effect is a large difference in the number of transverse modes in Andreev bound states. The oscillatory behavior of the supercurrent as a function of the length and chemical potential of the junction also differs substantially depending on the orientation of the superconducting electrodes deposited on the phosphorene sheet. For Andreev reflection, we show that gate voltaging controls the probability of this process and that the anisotropic behavior found in the supercurrent case is also present for conductance spectra.
\end{abstract}

\maketitle

%-------------------------------------------------------------------------------%
%                                     INTRO                                     %
%-------------------------------------------------------------------------------%
\section{Introduction}
There is currently much research addressing the physics of two-dimensional materials consisting of a single or very few atomic layers. Besides their interest from a fundamental viewpoint, such materials typically also feature unusual electronic properties which has spurred efforts to identify possible technological applications. Notable two-dimensional systems where the electrons display Dirac physics despite moving at non-relativistic velocities include graphene \cite{castroneto_rmp_09}, silicene \cite{kara_ssr_12}, and transition metal dichalcogenides \cite{duan_csr_15}. Phosphorene is a two-dimensional material which may be regarded as a single layer of black phosphorous, analogously to how graphene is a single layer of graphite. It was recently isolated by mechanical exfoliation \cite{li_natnano_14, koenig_apl_14, liu_acsnano_14} and has attracted attention due to the existence of an intrinsic band gap, in contrast to what occurs in graphene. This thickness-dependent band gap substantially increases the potential for using phosphorene in semiconductor-based technology such as transistors and solar cells. In addition to its promising electronic properties, phosphorene also displays high mechanical flexibility.

The proximity effect in two-dimensional materials was studied early on in graphene and various features of superconducting graphene have been clarified \cite{beenakker_prl_06, sengupta_prl_06, ludwig_prb_07, linder_prl_07} (more recently also in silicene \cite{linder_prb_14, li_prb_16, li_prb_16b, zhou_prb_16, li_prb_16c, kuzmanovski_prb_16, paul_prb_16}).
It has been shown that Andreev reflection in superconducting graphene junction is specular\cite{beenakker_prl_06}. Josephson current in ballistic graphene at the Dirac point is formally identical to that in a disordered normal metal\cite{Titov}.
The tunneling conductance\cite{sengupta_prl_06,linder_prl_07} and Josephson current\cite{Maiti} in graphene junctions are oscillatory functions of a width and height of the barrier at the interface.

In this paper, we study how superconducting correlations are manifested in phosphorene. In practice, this is accomplished via the proximity effect to a host superconducting material where a tunnel coupling allows Cooper pairs to penetrate a finite distance into the phosphorene sheet. We address two of the most fundamental superconducting transport phenomena, the Josephson effect \cite{josephson_pl_62} and Andreev reflection \cite{andreev_jetp_64, btk}, in a phosphorene sheet. In contrast to superconducting transport in \eg graphene and silicene, the supercurrent and conductance spectra for phosphorene are strongly anisotropic depending on how the superconducting electrodes are placed on the phosphorene sheet, differing with an order of magnitude depending on the orientation. Moreover, the supercurrent and conductance display oscillations as a function of the distance between the superconducting/normal electrodes with a period that also depends on the orientation of the electrodes. These results highlight that there may exists interesting opportunities with regard to tailoring anisotropic superconducting transport due to the geometry of the setup by using phosphorene.

%-------------------------------------------------------------------------------%
%                                     THEORY                                  %
%-------------------------------------------------------------------------------%
\section{Theory}

We here describe the derivation of the Andreev bound state (ABS) energies and the resulting supercurrent transport. The starting point is the two-band model of phosphorene \cite{ezawa_njp_14, sarkar_arxiv_16}
\begin{align}
H &= \sum_\vk \psi_\vk^\dag H_0(\vk) \psi_\vk,\notag\\
H_0(\vk) &= \begin{pmatrix}
f_\vk & g_\vk - \i h_\vk \\
g_\vk + \i h_\vk & f_\vk \\
\end{pmatrix}
\end{align}
where we use a basis vector of operators
\begin{align}
\psi_\vk^\dag = [c_{\vk,1}^\dag, c_{\vk,2}^\dag]
\end{align}
and the subscript $i$ on the fermionic $c_{\vk,i}^\dag$ creation operators denote the two atoms in the reduced unit cell. The quantities $f,g,h$ are defined as \cite{ezawa_njp_14, sarkar_arxiv_16}:
\begin{align}
f_\vk &= 4t_4\cos\Big(\frac{\sqrt{3}k_x}{2}\Big)\cos\Big(\frac{k_y}{2}\Big),\notag\\
g_\vk &= 2t_1\cos\Big(\frac{k_x}{2\sqrt{3}}\Big)\cos\Big(\frac{k_y}{2}\Big) + t_2\cos\Big(\frac{k_x}{\sqrt{3}}\Big) \notag\\
&+2t_3\cos\Big(\frac{5k_x}{2\sqrt{3}}\Big)\cos\Big(\frac{k_y}{2}\Big) + t_5\cos\Big(\frac{2k_x}{\sqrt{3}}\Big), \notag\\
h_\vk &= -2t_1\sin\Big(\frac{k_x}{2\sqrt{3}}\Big)\cos\Big(\frac{k_y}{2}\Big) + t_2\sin\Big(\frac{k_x}{\sqrt{3}}\Big) \notag\\
&+2t_3\sin\Big(\frac{5k_x}{2\sqrt{3}}\Big)\cos\Big(\frac{k_y}{2}\Big) - t_5\sin\Big(\frac{2k_x}{\sqrt{3}}\Big).
\end{align}
We have not shown the lattice constants $a_x$ and $a_y$ explicitly for brevity of notation. The normal-state eigenvalues are 
\begin{align}
\epsilon_{\pm,\vk} = f_\vk \pm \sqrt{g_\vk^2 + h_\vk^2},
\end{align}
where $\epsilon_{+,\vk}$ has a minimum at the $\Gamma$ point $\vk = (0,0)$. The band gap is 
\begin{align}
E_\text{gap} = 2\sqrt{g_\vk^2 + h_\vk^2}\Big|_{\vk=(0,0)} \simeq 1.5 \text{ eV}
\end{align}
which is much larger than the superconducting gap. Therefore, specular Andreev reflection is disregarded and we focus on the conduction band $\epsilon_{+,\vk}$. Performing a low-energy expansion around the $\Gamma$ point yields:
\begin{align}
f_\vk &\simeq t_4(4-3 k_x^2/2 - k_y^2/2) -\mu,\notag\\
g_\vk &\simeq m + \alpha k_x^2 + \beta k_y^2,\notag\\
h_\vk &\simeq \gamma k_x.
\end{align}
Above, $t_4<0$ and all the other parameters are defined via the hopping terms $t_i$ between various atoms \cite{ezawa_njp_14, sarkar_arxiv_16} (all in units of eV):
\begin{align}
t_1 &= -1.22,\; t_2 = 3.665,\; t_3 = -0.205,\; t_4 = -0.105, \notag\\
t_5 &= -0.055,\; m = 0.76,\; \gamma = 2.29,\; \alpha=-0.045,\; \beta=0.36.
\end{align}
Inserting these low-energy expansions into the expression for the normal-state eigenvalues, one obtains that the spectrum is parabolic in $k_y$ when $k_x=0$ whereas it is close to linear (but not strictly linear) in $k_x$ at $k_y=0$ since $\gamma\gg |\alpha|$ and $\gamma \gg |t_4|$. To describe proximity-induced superconductivity, we now add a standard superconducting term:
\begin{align}
H_\text{sc} = \sum_\vk (\Delta c_{1\vk}^\dag c_{1,-\vk}^\dag + \Delta c_{2\vk}^\dag c_{2,-\vk}^\dag + \text{h.c.}).
\end{align}\\
Any coupling between the conduction and valence bands due to Andreev reflection is irrelevant due to the large band gap $E_\text{gap}$, and we therefore simply project out the fermion operators belonging to the $\epsilon_{-,\vk}$ valence band. This can be done after rewriting the original fermion operators $c_{i,\vk}$ in terms of the band-basis operators which we denote $\eta_{\pm,\vk}$. We find (dropping momentarily the $\vk$ index for brevity)
\begin{align}
\begin{pmatrix}
\eta_+ \\
\eta_- \\
\end{pmatrix}
 = \frac{1}{\sqrt{2}} \begin{pmatrix}
1 & \frac{g-\i h}{R} \\
1 & \frac{-g + \i h}{R}\\
\end{pmatrix}
\begin{pmatrix}
c_1 \\
c_2 \\
\end{pmatrix}
\end{align}
where $R \equiv \sqrt{g^2+h^2}$. Inserting this into the superconducting Hamiltonian, using that $f$ and $g$ are symmetric in $\vk$ whereas $h$ is antisymmetric, and discarding all final terms that contain $\eta_-$, we end up with:
\begin{align}
H = \sum_\vk \epsilon_{+,\vk} \eta_{+,\vk}^\dag \eta_{+,\vk} +  \sum_\vk (\Delta \eta_{+,\vk}^\dag \eta_{+,-\vk}^\dag + \text{h.c.}).
\end{align}
In other words, the effective model is comprised of the Bardeen-Cooper-Schrieffer Hamiltonian, but with a new, anisotropic normal-state dispersion $\epsilon_{+,\vk}$. In the band basis, the order parameter remains of the conventional $s$-wave type. \\

\begin{figure}[b!]
\includegraphics[scale=0.49]{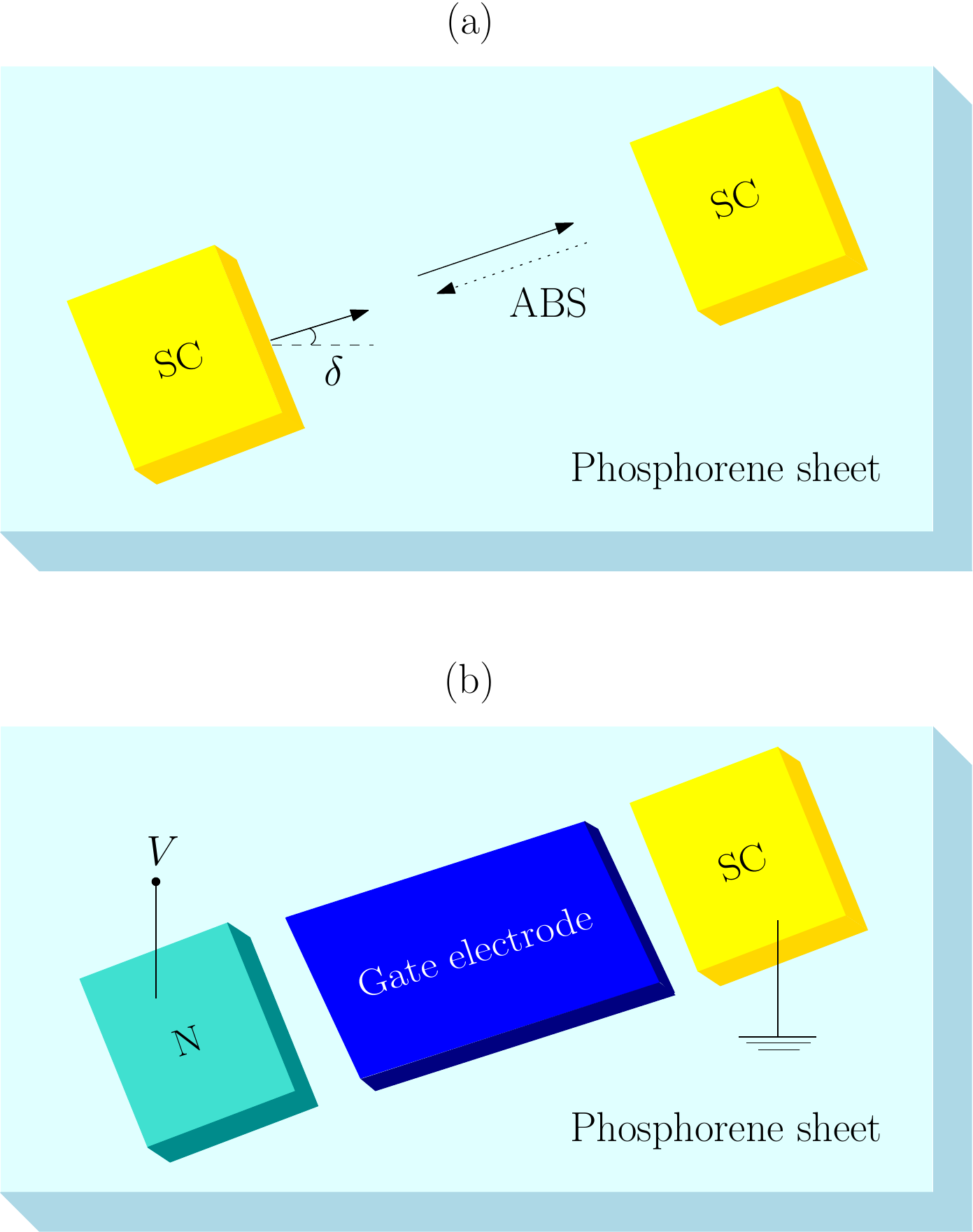}
\caption{(Color online) Top view of the proposed experimental setup for transport measurements. (a) Supercurrent measurements: two superconducting electrodes are deposited on top of a phosphorene sheet. A supercurrent flows between them upon current-biasing the system. The magnitude of the supercurrent depends on how the separation vector between the electrodes is oriented on the phosphorene sheet. This orientation is quantified by the angle $\delta$. (b) For conductance spectroscopy,  one of the superconducting electrodes is replaced with a normal metal electrode. A gate voltage between the electrodes can be used to tune the local chemical potential.}
\label{fig:model}
\end{figure}

\begin{figure}[t!]
\includegraphics[scale=0.55]{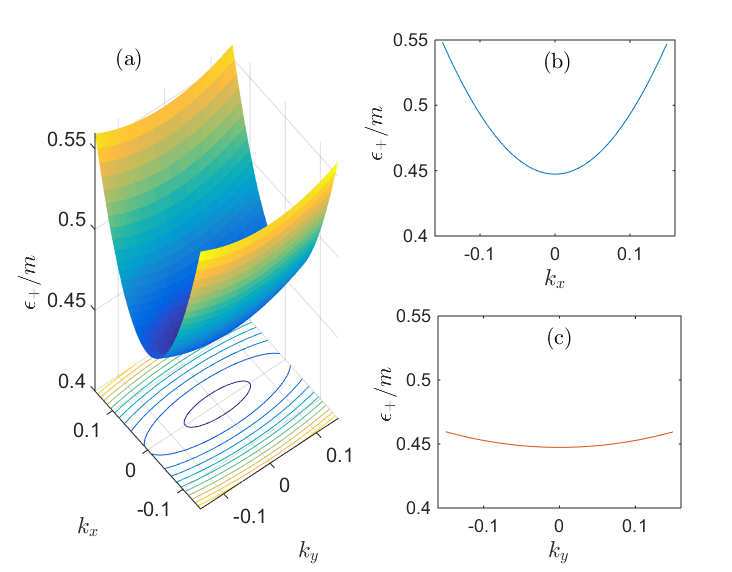}
\caption{(Color online) (a) Contour-plot of the upper band $\epsilon_+$ in the long-wavelength limit $(k_ja_j\ll1)$. Band-structure for $\epsilon_+$ for fixed (b)  $k_y=0$ and (c) $k_x=0$, demonstrating the anisotropy of the dispersion relation.}
\label{fig:fs}
\end{figure}

We will consider the formation of ABS in an superconductor/normal/superconductor (SNS) junction \textit{which does not necessarily extend along the $x$-axis}. In this way, we will be able to probe the effect of the anisotropic band-structure of phosphorene on Andreev reflection and the supercurrent. We use an extended version of the Blonder-Tinkham-Klapwijk formalism \cite{btk} adapted to materials  with a dispersion relation deviating from that of free electrons, using a similar procedure and notation as in \cite{linder_prl_08, linder_prb_14}. Let $\hat{n}$ be the interface normal to the SN interfaces. Usually, one considers $\hat{n} = \hat{x}$. Now, we want to consider any $\hat{n}$ between $\hat{x}$ and $\hat{y}$ to probe how the supercurrent changes due to the anisotropic normal-state dispersion which is linear in momentum in one direction and quadratic in momentum in another. The setup is shown in Fig. \ref{fig:model} where the angle $\delta$ defines the orientation of the electrodes, so that $\hat{n} = (\cos\delta,\sin\delta)$. For a given orientation, transverse modes will be fully taken into account. The dispersion of the upper band $\epsilon_+$ is shown in Fig. \ref{fig:fs}.

Realistically, there will be a Fermi vector-mismatch between the S and N regions due to charge-transfer between the SC electrode and the phosphorene layer under it. Thus, our calculations include the possibility for different chemical potentials $\mu$ in the two regions, i.e. $\mu_S$ and $\mu_N$. We will consider doped phosphorene where the chemical potential lies in the conduction band, allowing for a finite density of states that thus supports the formation of a proximity-induced superconducting state. The resulting boundary conditions at each interface of the structures that we consider, both in the supercurrent setup shown in Fig. \ref{fig:model}(a) and the conductance spectroscopy setup shown in (b), are:
\begin{align}
\psi_\text{left} = \psi_\text{right},\; \hat{v}_n \psi_\text{left} = \hat{v}_n \psi_\text{right}
\end{align}
where $\psi_\text{left/right}$ refers to the wavefunction on the left/right side of a given interface. The velocity operator is defined as $\hat{v}_n = \partial \mathcal{H}/\partial k_n$ and
\begin{align}
\mathcal{H} = 
\begin{pmatrix}
\epsilon_{+,\vk} & \Delta\\
\Delta^* & -\epsilon_{+,\vk} \\
\end{pmatrix}
\end{align}
where the normal-state dispersion may be expressed via the momentum parallell to $\hat{n}$ and perpendicularly to it (instead of $k_x$ and $k_y$) by using that
\begin{align}
\begin{pmatrix}
k_x\\
k_y \\
\end{pmatrix}
 = \begin{pmatrix}
\cos\delta & - \sin\delta \\
\sin\delta & \cos\delta \\
\end{pmatrix}
\begin{pmatrix}
k_n\\
k_\perp \\
\end{pmatrix}.
\end{align}
Note that when applying the velocity operator, care should be taken to use the correct value of the momentum perpendicular to the interface which is determined by the chemical potential $\mu_S$ or $\mu_N$ (the S or N region). Let $c\equiv\cos\delta$ and $s\equiv\sin\delta$. In this case, we get 
\begin{align}
\epsilon_{\vk} \equiv \epsilon_{+,\vk} = f_\vk + \sqrt{g_\vk^2+ h_\vk^2}
\end{align}
where we defined
\begin{align}
f_\vk &= t_4[4-\frac{3}{2}(k_nc - k_\perp s)^2 - \frac{1}{2}(k_ns + ck_\perp)^2]-\mu,\notag\\
g_\vk &= m + \alpha(k_n c - k_\perp s)^2 + \beta(k_n s + ck_\perp)^2,\notag\\
h_\vk &= \gamma(k_nc - k_\perp s).
\end{align}

To determine the properties of the Josephson effect in an SNS phosphorene junction, we set up the wavefunctions in each region (left S, normal, right S):
\begin{align}
\psi_L &= \Bigg[ L_e\begin{pmatrix}
1 \\
\e{\i\zeta}\\
\end{pmatrix}\e{-\i k_nn} + L_h \begin{pmatrix}
\e{\i\zeta}\\
1\\
\end{pmatrix}\e{\i k_nn}
\Bigg] \e{\i k_\perp n_\perp},\notag\\
\psi_N &= \Bigg[ a\begin{pmatrix}
1 \\
0 \\
\end{pmatrix}
\e{\i k_n' n} + b\begin{pmatrix}
1 \\
0 \\
\end{pmatrix}
\e{-\i k_n' n}  + c\begin{pmatrix}
0 \\
1 \\
\end{pmatrix}
\e{\i k_n' n}  + d\begin{pmatrix}
0 \\
1 \\
\end{pmatrix}
\e{-\i k_n' n} 
\Bigg] \e{\i k_\perp n_\perp} \notag\\
\psi_R &= \Bigg[ R_e\begin{pmatrix}
1 \\
\e{\i\zeta-\i\phi}\\
\end{pmatrix}\e{\i k_nn} + R_h \begin{pmatrix}
\e{\i\zeta+\i\phi}\\
1\\
\end{pmatrix}\e{-\i k_nn}
\Bigg] \e{\i k_\perp n_\perp},
\end{align}
Here, we have allowed for a different $\mu$ in the N and S regions by distinguishing the wavevectors $k_n'$ and $k_n$ in these regions while $n$ and $n_\perp$ denote the coordinates parallell to and perpendicular to the interface normal $\hat{n}$. Moreover, $\phi$ is the SC phase difference and $\zeta=\text{acos}(E/\Delta_0)$ where $E$ is the quasiparticle energy. The wavevectors $\{k_n,k_n'\}$ have to be obtained from the dispersion relation numerically for a fixed value of $k_\perp$. For a given direction of the electrodes $\hat{n}$, we only consider the contribution from propagating modes in the N region, i.e. only contributions from the $k_\perp$ values that give a real $k_n$ and $k_n'$. This is required to be consistent with the diagonalization of the Hamiltonian. Unlike Ref. \cite{sarkar_arxiv_16}, we will not make any approximations in the band-structure (such as setting $\alpha=0$) and we thus keep all terms in the dispersion relation.\\

Defining the quantity 
\begin{align}
F(k) \equiv \frac{\partial \epsilon_\vk}{\partial k_n} \Bigg|_{k_n=k},
\end{align}
we can write down the system of equations that determine the ABS energies:
\begin{align}
L_e + L_h \e{\i\zeta} &= a+b,\notag\\
L_e\e{\i\zeta} + L_h &= c+d,\notag\\
R_e \e{\i k_nL} + R_h\e{\i\zeta+\i\phi}\e{-\i k_nL} &= a\e{\i k_n'L} + b\e{-\i k_n'L},\notag\\
R_e\e{\i\zeta-\i\phi}\e{\i k_nL} + R_h \e{-\i k_nL} &= c\e{\i k_n'L} + d\e{-\i k_n'L},
\end{align}
stem from the continuity of the wavefunction, whereas
\begin{align}
F(-k_n)L_e + F(k_n) L_h\e{\i\zeta} &= F(k_n')a + F(-k_n')b,\notag\\
F(-k_n)L_e\e{\i\zeta}+ F(k_n)L_h &= F(k_n')c + F(-k_n')d,
\end{align}
and also
\begin{align}
F(k_n)R_e\e{\i k_nL} &+ F(-k_n)R_h\e{\i\zeta+\i\phi}\e{-\i k_nL} \notag\\
 = F(k_n')&\e{\i k_n'L}a + F(-k_n')\e{-\i k_n'L}b,\notag\\
F(k_n)R_e\e{\i k_nL}\e{\i\zeta-\i\phi} &+ R_h F(-k_n)\e{-\i k_nL} \notag\\
 = F(k_n')&c\e{\i k_n'L} + F(-k_n')d\e{-\i k_n'L}.
\end{align}
stem from the continuity of particle flux. By writing this system of equations as $Ax=0$ where $A$ is an $8\times8$ matrix and $x$ is a vector containing all the scattering coefficients, one determines the ABS energies from the requirement det($A$)=0. The supercurrent is then obtained from:
\begin{align}\label{eq:supercurrent}
I = \frac{2e}{\hbar} \sum_{k_\perp} \sum_\pm \frac{\partial E_\pm}{\partial\phi} f(E_\pm),
\end{align}
where the ABS energies have the form $E_\pm = \pm \Delta_0\sqrt{R(\phi)}$. Here, $f$ is the Fermi-Dirac distribution function. This expression is derived from the fundamental thermodynamical relation between the free energy $F$ of the junction and the supercurrent $I$, namely $(2e/\hbar)dF/d\phi=I$. It is valid \cite{beenakker_prl_91} in the short-junction limit $L/\xi\ll 1$ which we will consider in this paper. Here,$L$ and $\xi$ are the length of the junction and superconducting coherence length, respectively.  The expression for $R(\phi)$ is too lengthy to be particularly useful, but may be numerically implemented. 
Using Eq. (\ref{eq:supercurrent}), we can then study how the supercurrent differs for propagation along the $x$-axis ($\delta=0$) and the $y$-axis ($\delta=\pi/2$), as well as its dependence on the length $L$ of the junction. For the transverse modes, we use an effective width $W=500$ nm for the junction in both cases with a spacing of $\pi/W$ between the transverse modes, and the temperature is set to $\Delta_0/k_BT = 100$ (corresponding to $T\ll T_c$). Our choice of width $W\gg L$ is large enough to effectively mimic a semi-infinite junction in the transverse direction, as the results are quantitatively indistinguishable upon further increasing $W$.

To study Andreev reflection and conductance spectra in a phosphorene junction, a NN'S junction is described by the following wavefunctions:
\begin{align}
\psi_N &= \Bigg[ \begin{pmatrix}
1 \\
0 \\
\end{pmatrix}
\e{-\i k_n' n}  + a\begin{pmatrix}
0 \\
1 \\
\end{pmatrix}
\e{\i k_n' n} + b\begin{pmatrix}
1 \\
0 \\
\end{pmatrix}
\e{-\i k_n' n} 
\Bigg] \e{\i k_\perp n_\perp} \notag\\
\psi_{N'} &= \Bigg[
c\begin{pmatrix}
1 \\
0 \\
\end{pmatrix}
\e{\i k_n'' n} 
+ d\begin{pmatrix}
1 \\
0 \\
\end{pmatrix}
\e{-\i k_n'' n} +e \begin{pmatrix}
0 \\
1 \\
\end{pmatrix}
\e{\i k_n'' n}  + f\begin{pmatrix}
0 \\
1 \\
\end{pmatrix}
\e{-\i k_n'' n} 
\Bigg] \e{\i k_\perp n_\perp} \notag\\
\psi_S &= \Bigg[ R_e\begin{pmatrix}
1 \\
\e{\i\zeta-\i\phi}\\
\end{pmatrix}\e{\i k_nn} + R_h \begin{pmatrix}
\e{\i\zeta+\i\phi}\\
1\\
\end{pmatrix}\e{-\i k_nn}
\Bigg] \e{\i k_\perp n_\perp}.
\end{align}
With the boundary conditions, we obtain the scattering coefficients. The expressions for the coefficients $a$ and $b$ are given in the Appendix.

\begin{figure}[t!]
\begin{centering}
\includegraphics[width=0.5\textwidth]{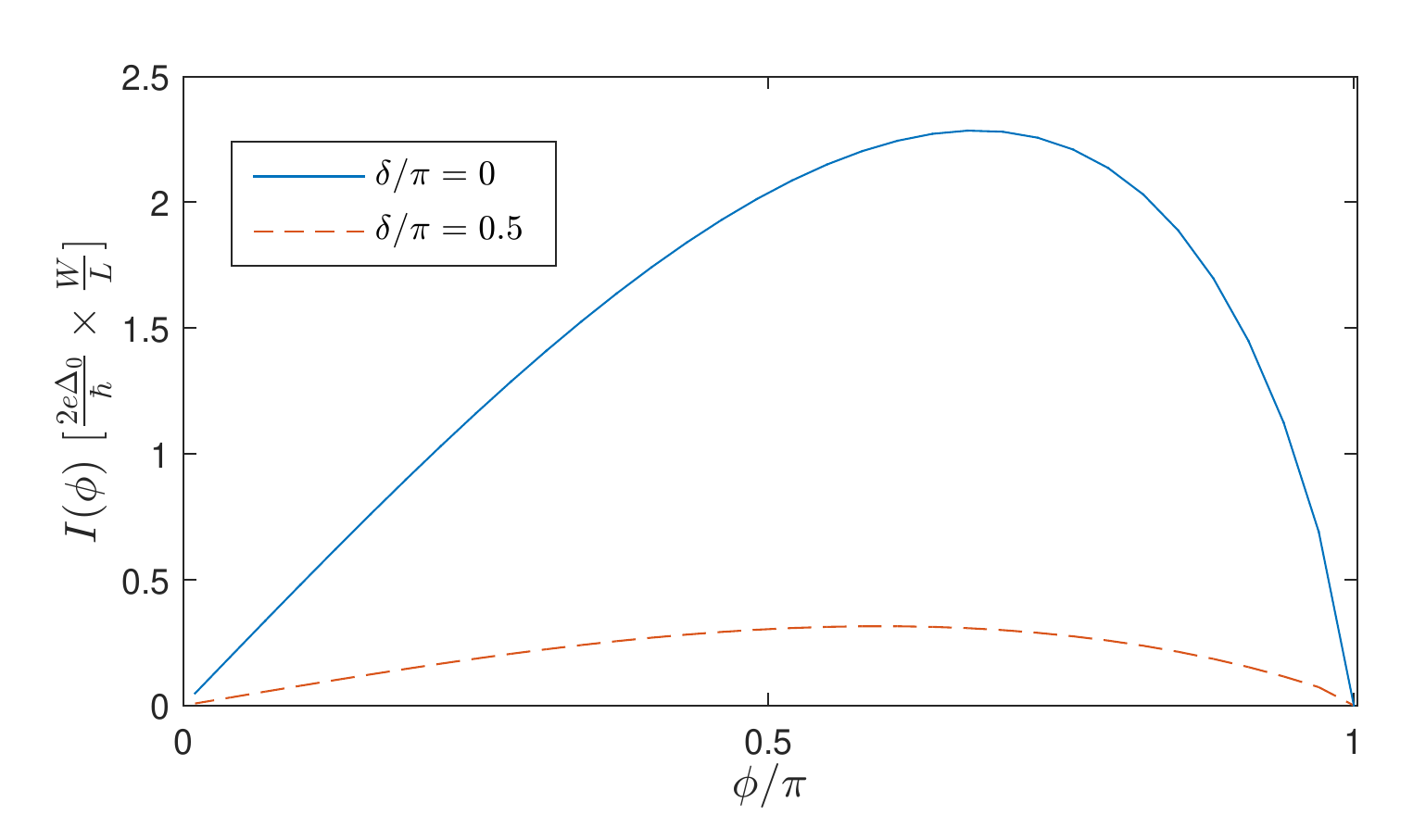}
\caption{(Color online) Supercurrent-phase relation for $\mu_N=4t_4 + 1.05m$ for propagation directions $\delta=0$ and $\delta=\pi/2$. The magnitude of the critical current changes by nearly an order of magnitude. }
\label{fig:delta}
\end{centering}
\end{figure}

The normalized conductance at zero temperature is then calculated as 
\begin{align}
\sigma  = \frac{{\sum\limits_{{k_ \bot }} {F({k_n'})\left( {1 + {{\left| a \right|}^2} - {{\left| b \right|}^2}} \right)} }}{{\sum\limits_{{k_ \bot }} {F({k_n'})} }}.
\end{align}
We have included a normal region N' separating the N and S regions in order to study how the transport properties are affected by gating on N', which controls the local chemical potential.

%-------------------------------------------------------------------------------%
%                         RESULTS AND DISCUSSION                   %
%-------------------------------------------------------------------------------%
\section{Results and Discussion}

\subsection{Josephson effect}
We begin by considering the direction dependence for a fixed length $L=15$ nm, corresponding to $L/W=0.03$. Due to the charge-transfer between the host superconductor and the region of the phosphorene sheet directly underneath it, we set $\mu_S>\mu_N$ in order to use experimentally relevant values. Specifically, we set $\mu_S = 4t_4 + 1.5m$. As a reference value, we note that $\mu=4t_4+m$ corresponds to the bottom of the conduction band. Fig. \ref{fig:delta} shows the supercurrent vs. phase. It is clear that the supercurrent is highly anisotropic and differs by an order of magnitude when comparing propagation in the $x$- and $y$-directions ($\delta=0$ and $\delta=\pi/2$, respectively) for \eg $\mu_N = 4t_4 + 1.05m$ (meaning that the N region is assumed to be only slightly doped). To understand this, one should note that not only the effective dispersion relation is different along these directions, but the number of allowed propagating modes also differs greatly due to the anisotropic Fermi surface. As a result, the supercurrent magnitude becomes larger in the $x$-direction along which the dispersion is closer to being linear, hosting more propagating modes characterized by $k_\perp$. The maximum of the supercurrent is shifted slightly away from $\phi=\pi/2$ due to the presence of higher harmonics which are typically present in high-transparency ballistic Josephson junctions.

\begin{figure}[t!]
\begin{centering}
\includegraphics[width=0.5\textwidth]{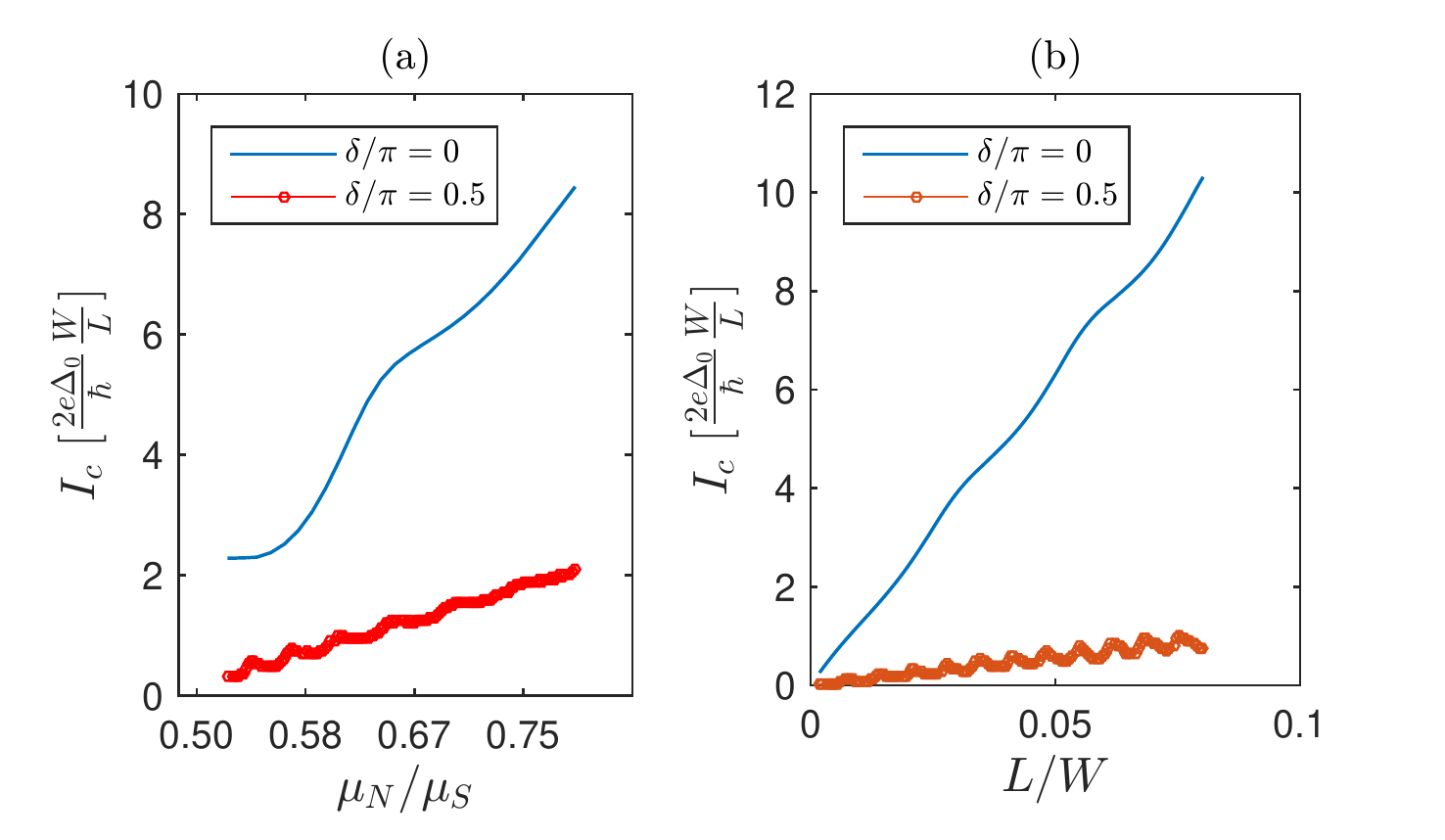}
\caption{(Color online) \textit{Left panel}: Critical current vs. the chemical potential for $L/W=0.03$. \textit{Right panel}: Critical current vs. the length $L$ of the junction for $\mu_N=4t_4 + 1.05m$.  }
\label{fig:muL}
\end{centering}
\end{figure}

Next, we consider the length and chemical potential dependence of the supercurrent. The supercurrent shows oscillation due to Klein tunneling pertaining to Dirac fermions.\cite{Maiti} 
Besides the difference in magnitude, a qualitative difference emerges between the two directions. In the $\delta=\pi/2$ direction, the oscillation period is much smaller than in the $\delta=0$ direction. This is physically reasonable upon considering the different dispersion relations which affects the wavevector magnitude and thus the oscillation period, since the supercurrent depends on the product $k_nL$. For a fixed value of the Fermi level $\mu_N$, the Fermi wavevector is much smaller in the $\delta=0$ case for normal incidence $k_y=0$ in Fig. \ref{fig:fs} than the Fermi wavevector in the $\delta=\pi/2$ case for normal incidence $k_x=0$. As a result, the oscillations occur on a shorter length-scale in the $\delta=\pi/2$ case as shown in Fig. \ref{fig:muL}(b).

\subsection{Andreev reflection}
We next consider how Andreev reflection is manifested in phosphorene. Experimentally, gating is a commonly used way to influence the transport properties. For this reason, we take into account a phosphorene region covered by a gate electrode that separates the normal and superconducting parts. In effect, we are considering an NN'S phosphorene junction where the chemical potentials are denoted by $\mu_N$, $\mu_{N'}$, and $\mu_S$. In the superconducting region, we fix $\mu_S=4t_4 + 1.5m$, and we proceed to determine how the conductance spectra of the system under consideration depends on applied bias voltage $eV$ and the chemical potentials $\mu_N$ and $\mu_{N'}$. We fix the length $L$ of the N' layer to $L=$10 nm. 

\begin{figure}[t!]
\begin{centering}
\includegraphics[width=0.5\textwidth]{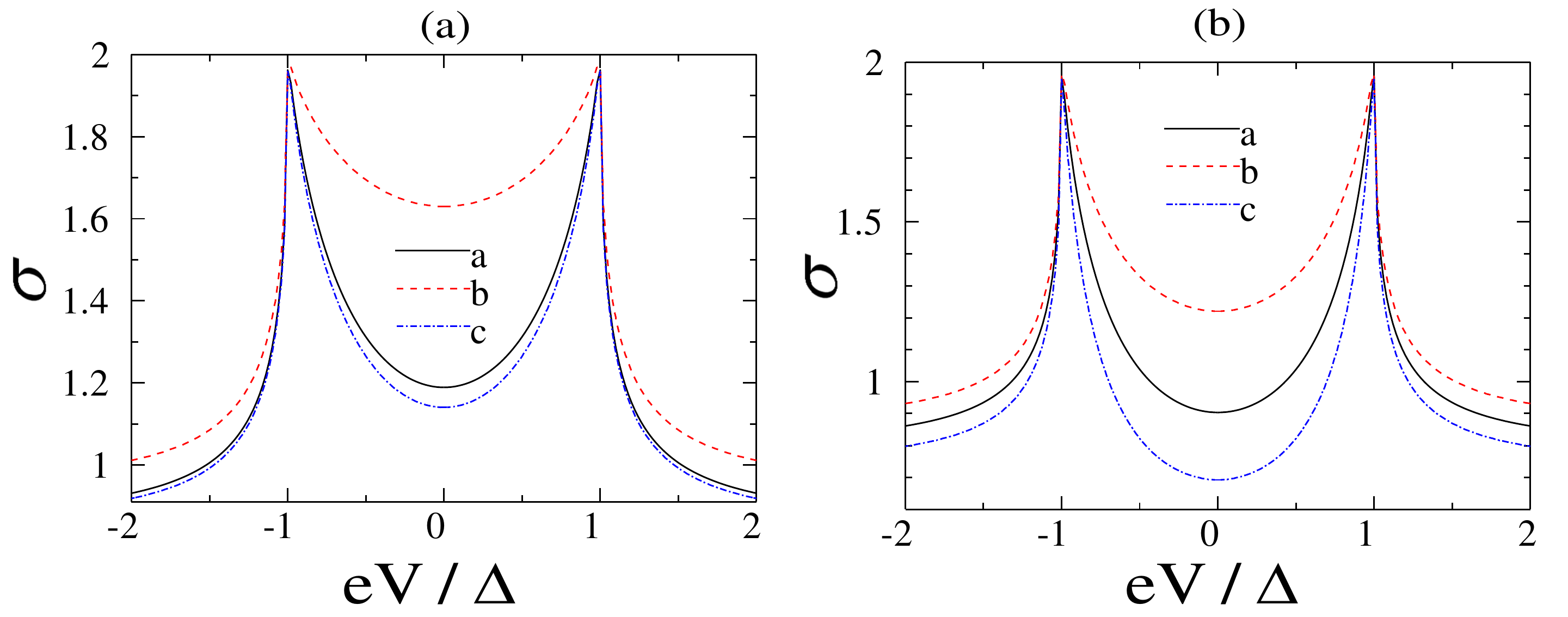}
\caption{(Color online) Normalized conductance for $\mu_N=4t_4 + 1.01m$ for (a) $\delta=0$ and (b) $\delta=\pi/2$. The lines in the panels correspond to: a. $\mu_{N'}=4t_4 + 1.1m$, b. $\mu_{N'}=4t_4 + 1.3m$, c. $\mu_{N'}=4t_4 + 1.6m$. }
\label{fig2}
\end{centering}
\end{figure}

\begin{figure}[t!]
\begin{centering}
\includegraphics[width=0.5\textwidth]{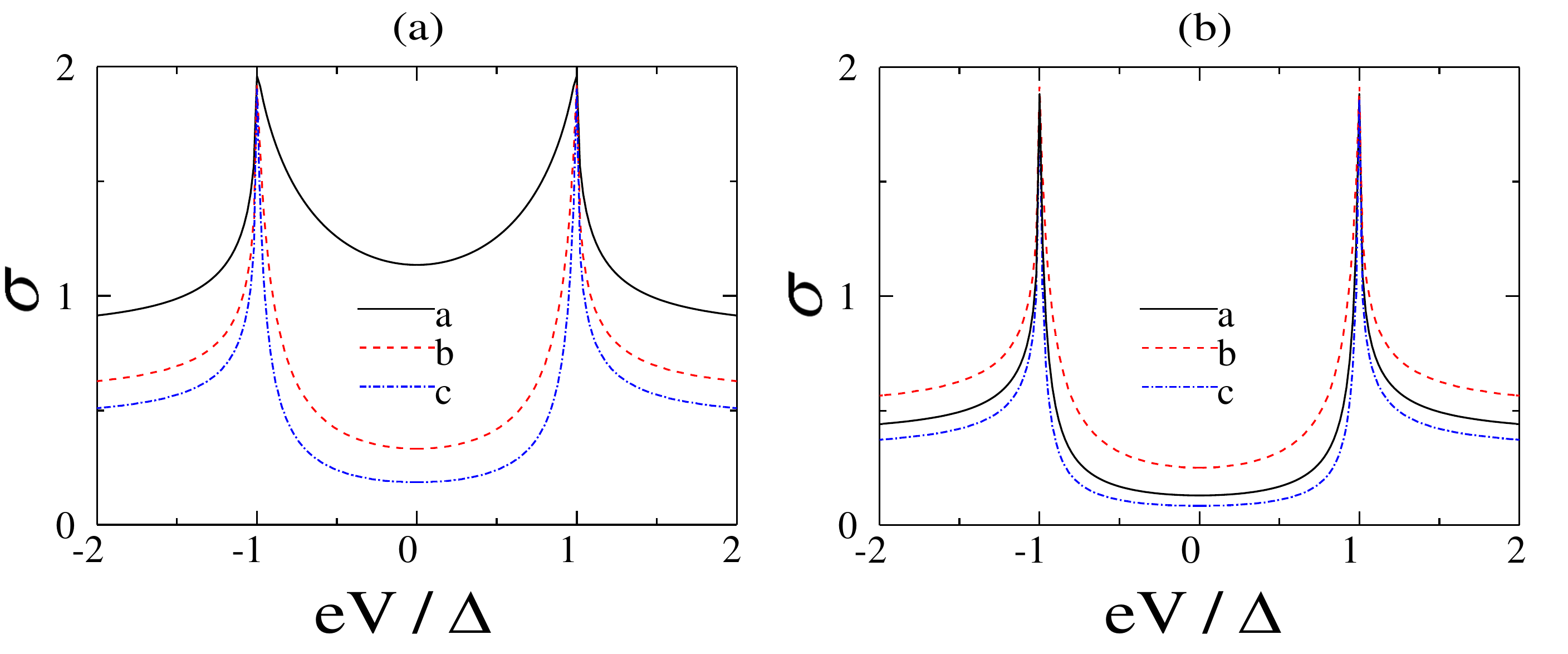}
\caption{(Color online) Normalized conductance for $\mu_N=4t_4 + 1.1m$ for (a) $\delta=0$ and (b) $\delta=\pi/2$. The lines in the panels correspond to: a. $\mu_{N'}=4t_4 + 1.1m$, b. $\mu_{N'}=4t_4 + 1.3m$, c. $\mu_{N'}=4t_4 + 1.6m$.}
\label{fig1}
\end{centering}
\end{figure}

\begin{figure}[t!]
\begin{centering}
\includegraphics[width=0.5\textwidth]{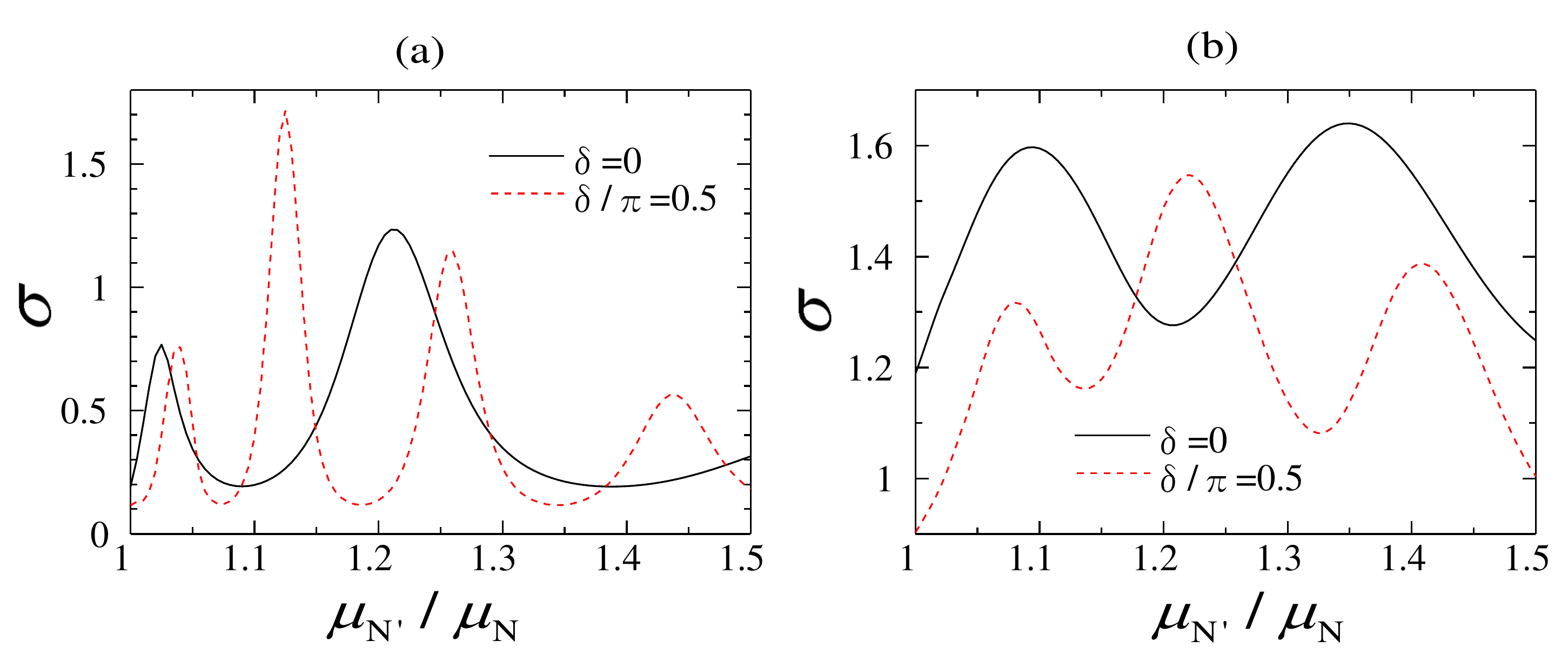}
\caption{(Color online) Normalized conductance at zero bias as a function of $\mu_{N'}$ for (a) $\mu_N=4t_4 + 1.01m$ and for (b) $\mu_N=4t_4 + 1.1m$. }
\label{fig3}
\end{centering}
\end{figure}

In Fig. \ref{fig2}, we show the conductance for $\mu_N=4t_4 + 1.01m$ with (a) $\delta=0$ and (b) $\delta=\pi/2$ and several values of $\mu_{N'}$. A clear anisotropy is seen for the two directions $\delta=0$ and $\delta=\pi/2$. Due to a large Fermi wave-vector mismatch, Andreev reflection is suppressed and the conductance shows a gap-like structure. Figure \ref{fig1} depicts the conductance for $\mu_N=4t_4 + 1.1m$ with (a) $\delta=0$ and (b) $\delta=\pi/2$ and several values of $\mu_{N'}$. The conductance obtained for the two directions $\delta=0$ and $\delta=\pi/2$ remains noticeably different. Since the Fermi wavevector mismatch is small, the gap-like structures become shallower. In Fig. \ref{fig3}, we show the conductance at zero bias as a function of $\mu_{N'}$ for (a) $\mu_N=4t_4 + 1.01m$ and (b) $\mu_N=4t_4 + 1.1m$. The conductance oscillates as a function of $\mu_{N'}$ with different periods for two directions. These oscillations stem from wavenature of the wavefunction for Dirac fermions in the N' region. As explained in the previous subsection, the oscillation period is smaller in the $\delta=\pi/2$ case due to the difference in Fermi wavevector magnitude when comparing the two directions of propagation. Moreover, the result indicates that by gating, one can tune the anisotropy of the Andreev reflection probability and conductance. Note that the non-normalized conductances for $\delta=0$ and $\pi/2$ differ in magnitude with about one order due to the different number of the transverse modes $k_\perp$ contributing to the transport.

%-------------------------------------------------------------------------------%
%                                   CONCLUSION                                  %
%-------------------------------------------------------------------------------%

\section{Conclusion}
In conclusion, we have presented a study of the anisotropic superconducting transport properties of phosphorene, a single layer of black phosphorous. The system setup consisting of superconducting or normal electrodes deposited at different locations of a phosphorene sheet should be experimentally feasible in light of the recent experimental reports of stable, isolated phosphorene via exfoliation. Due to the anisotropic band structure of this system, the supercurrent magnitude changes with an order of magnitude when comparing tunneling along two perpendicular directions in the monolayer. The oscillatory behavior of the supercurrent as a function of the length and chemical potential of the junction is different when modifying the orientation of the superconducting electrodes deposited on the phosphorene sheet. For Andreev reflection, we show that gate voltaging controls the probability of this process and that the anisotropic behavior found in the supercurrent case is also present for conductance spectra. 
The oscillatory behaviors of the supercurrent and conductance found here are manifestation of Dirac dispersions, as seen in graphene junctions.\cite{sengupta_prl_06,Maiti}

Interesting future directions to explore include non-local transport in multiterminal geometries and in particular the crossed Andreev reflection process, as well as the inclusion of magnetic elements in the system setup.

%-------------------------------------------------------------------------------%
%                                     CREDIT                                    %
%-------------------------------------------------------------------------------%
\acknowledgments
J.L. acknowledges funding via the Outstanding Academic
Fellows program at NTNU, the NV-Faculty, and the Research
Council of Norway Grant numbers 216700 and 240806.
This work was supported by a Grant-in-Aid for Scientific Research on Innovative Areas "Topological Materials Science" (KAKENHI Grant No. JP16H00988) from JSPS of Japan.

%-------------------------------------------------------------------------------%
%                                     Appendix                                   %
%------------------------------------------------------------------------------%
\begin{widetext}
\appendix

\section{Scattering coefficients}
The coefficients $a$ and $b$ are given by
\begin{align}
a = \frac{{uv\left[ {(A{\alpha ^{ - 1}} + B\alpha )(A\alpha  + B{\alpha ^{ - 1}}) - (C{\alpha ^{ - 1}} + D\alpha )(C\alpha  + D{\alpha ^{ - 1}})} \right]}}{{(A{\alpha ^{ - 1}} + B\alpha )(A\alpha  + B{\alpha ^{ - 1}}){u^2} - (C{\alpha ^{ - 1}} + D\alpha )(C\alpha  + D{\alpha ^{ - 1}}){v^2}}},\\
b = \frac{{(A\alpha  + B{\alpha ^{ - 1}})(C\alpha  + D{\alpha ^{ - 1}})({u^2} - {v^2})}}{{(A{\alpha ^{ - 1}} + B\alpha )(A\alpha  + B{\alpha ^{ - 1}}){u^2} - (C{\alpha ^{ - 1}} + D\alpha )(C\alpha  + D{\alpha ^{ - 1}}){v^2}}}
\end{align}
where 
\begin{align}
A = \left( {F({k_n}') + F({k_n}'')} \right)\left( {F({k_n}'') + F({k_n})} \right),\quad B = \left( {F({k_n}') - F({k_n}'')} \right)\left( {F({k_n}'') - F({k_n})} \right),\\
C = \left( {F({k_n}') + F({k_n}'')} \right)\left( {F({k_n}'') - F({k_n})} \right),\quad D = \left( {F({k_n}') - F({k_n}'')} \right)\left( {F({k_n}'') + F({k_n})} \right),\\
u = \sqrt {\frac{1}{2}\left( {1 + \frac{{\sqrt {{{(eV)}^2} - {\Delta ^2}} }}{{eV}}} \right)} ,\quad v = \sqrt {\frac{1}{2}\left( {1 - \frac{{\sqrt {{{(eV)}^2} - {\Delta ^2}} }}{{eV}}} \right)} 
\end{align}
and $\alpha  = {e^{i{k_n}''L}}$. 
\end{widetext}

%-------------------------------------------------------------------------------%
%                                  BIBLIOGRAPHY                                 %
%-------------------------------------------------------------------------------%

\end{document}